\UseRawInputEncoding
\documentclass[5p,times]{elsarticle}
\usepackage{latexsym}
\usepackage{amssymb,amsmath,amsfonts}
\usepackage{multirow}
\usepackage{multicol}
\usepackage{mathtools}
\numberwithin{equation}{section}
\journal{Partial Differential Equations in Applied Mathematics}
\newtheorem{example}{Example}[section]
\newtheorem{remark}{Remark}[section]
\begin{document}

\begin{frontmatter}

\title{Exact solutions of nonlinear delay reaction-diffusion equations with variable coefficients}
\author[a1,a2,a3]{M.O. Aibinu\corref{c1}}
\ead{moaibinu@yahoo.com / mathewa@dut.ac.za}
\author[a4]{S. C. Thakur}
\ead{thakur@dut.ac.za}
\author[a5]{S. Moyo}
\ead{dvcrie@dut.ac.za}
\address[a1]{Institute for Systems Science \& KZN E-Skill CoLab, Durban University of Technology, Durban 4000, South Africa}
\address[a2]{DSI-NRF Centre of Excellence in Mathematical and Statistical Sciences (CoE-MaSS), South Africa}
\address[a3]{National Institute for Theoretical and Computational Sciences (NITheCS), South Africa}
\address[a4]{KZN E-Skill CoLab, Durban University of  Technology, Durban 4000, South Africa}
\address[a5]{Institute for Systems Science \& Office of the DVC Research, Innovation \& Engagement\\
Milena Court, Durban University of Technology, Durban 4000, South Africa}
\cortext[c1]{Corresponding author}

\begin{abstract}
A modified method of functional constraints is used to construct the exact solutions of nonlinear equations of reaction-diffusion type with delay and which are associated with variable coefficients. This study considers a most generalized form of nonlinear equations of reaction-diffusion type with delay and which are nonlinear and associated with variable coefficients. A novel technique is used in this study to obtain the exact solutions which are new and are of the form of traveling-wave solutions. Arbitrary functions are present in the solutions and they also contain free parameters, which make them suitable for usage in solving certain modeling problems, testing numerical and approximate analytical methods. The results of this study also find applications in obtaining the exact solutions of other forms of partial differential equations which are more complex. Specific examples of nonlinear equations of reaction-diffusion type with delay are given and their exact solutions are presented. Solutions of certain reaction-diffusion equations are also displayed graphically.
\end{abstract}
\begin{keyword}
Reaction-diffusion\sep Time delay\sep Exact solutions\sep Differential equations.
\MSC[2010] 35K55 \sep 35B10 \sep 35B40 \sep 35K57

\end{keyword}
\end{frontmatter}

\section{Introduction}
The roles which Nonlinear Partial Differential Equations (NPDEs) play are prominent in the description and analysis of real-life processes and phenomena (See e.g., \cite{Nisar4, Nisar5}). Therefore, it is pivotal to seek for the ways of obtaining exact solutions of NPDEs for a proper and accurate analysis. Several processes and phenomena which occur in sciences and engineering lead to NPDEs as there are several conditions and parameters to be considered in the modeling of such systems. Reaction-Diffusion Equations (RDEs) are members of  NPDEs. Reaction-diffusion systems can be described as mathematical models which find applications in diverse physical phenomena. In its simplest form in one spatial dimension, the RDE has the form
 \begin{equation}\label{rd27}
u_t=Du_{xx}+G(u),
 \end{equation}
where $u(x,t)$ denotes the unknown function, $G$ accounts for all local reactions, and $D$ is a diffusion coefficient (which is a constant) (See e.g, \cite{Kolmogorov}). RDEs are pervading in the mathematical modeling of the systems which occur in biology, chemistry, complex physics phenomena, engineering, and mechanics \cite{Murray}. RDEs can also represent the chemical reactions and diffusion processes. Basically many real-life processes do not only depend on the present state but also past occurrences. Also, the dynamical systems are constituted by the time delay. The study of nonlinear delay RDEs provides a fundamental tool for quantitative and qualitative analyses of various dynamical systems such as those modeling infections. For RDEs with delay,  the kinetic function $G$ which denotes the chemical reactions rates is a function of both $u = u(x, t)$ and $w = u(x, t-\tau),$ which represent the sought after concentration function and delayed argument, respectively. Two special cases which can arise are $G(u, w) = g(w)$ and $G(u, w) = g(u).$ A system with local non-equilibrium media is described by $G(u, w) = g(w).$ These are systems which possess inertial properties and reactions will always begin after a time $\tau.$ $G(u, w) = g(u)$ represents the classical local equilibrium case (See e.g., \cite{Polyanin6}).
\par  How to obtain the solutions of RDEs has recently attracted much attention. NPDEs are universal in nature and for finding their solutions, several methods have been employed which include spectral collocation and waveform relaxation \cite{Jackiewicza, Kumbinarasaiah, Aziz}, adomian decomposition \cite{Alderremy, Elzaki}, Tan-Cot \cite{Al-Shaeer}, residual power series \cite{Rajarama}, perturbation \cite{Bilidik, Jena}.  However, there are disadvantages which are commonly associated with those methods. There are conditions which make the universal application of those listed methods and others to be impossible. The objects are different in their geometric shapes. The reaction kinetics and type of fluid flow are erratic. The worthlessness in the presence of singular points is indisputable. Also, efforts have been made on using certain recently proposed numerical and iterative methods, which are widely remarked for their accuracy and effectiveness in obtaining the solutions of high-dimensional complex geometry nonlinear problems (See e.g., \cite{Wang1, Wang2, Nisar1, Nisar2, Nisar3, Nisar6, Aibinu}).  Obtaining the exact solutions is imperative for proper analysis of the processes which are under consideration (localization, nonuniqueness, blowup regimes, spatial, etc). 
\par Subsequently, the term "exact solution" in relating to NPDEs will refer to where the solution can be expressed in:
\begin{itemize}
	\item [(i)]terms of elementary functions;
	\item [(ii)]closed form with definite or/and indefinite integrals;
	\item [(iii)]
	terms of solutions to Ordinary Differential Equations (ODEs) or systems of such equations.
\end{itemize}
Accepted form for exact solutions also includes the combinations of cases listed above (See e.g, \cite{Polyanin6, Polyanin1, Polyanin2}). A most general form of (\ref{rd27}) is when nonlinear terms are present before the time derivative and the diffusion term. An awkward form occurs when the combination of the kinetic function $G$ and nonlinear terms are also associated with the variable derivative.  This present study is motivated by how to construct the exact solutions for a most general form of nonlinear RDEs which are awkward in nature.

\par Let $G(u, w)$ be an arbitrary function which takes two arguments, $\tau > 0$ be the delay time, while $a(x), b(x), c(x), p(x)$ and $q(x)$ are some functions. This study considers nonlinear delay RDEs of the form
 \begin{equation}\label{rd1}
 \begin{split}
&\left[q(x)+c(x)g(u)\right]u_t=\left[a(x)u_x\right]_x\\
&+\left[p(x)+b(x)G\left(u, w\right)\right]u_x, ~~w=u(x, t-\tau).
\end{split}
 \end{equation}   
 This is a most generalized form with the presence of variable coefficients and arbitrary functions. The exact solutions are obtained in the form of generalized  traveling-wave solutions. A novel technique is presented for constructing the exact solutions of a most generalized form of nonlinear RDEs with variable coefficients and with delay. The exact solutions for coupled reaction-diffusion nonlinear system with delay are also obtained. The technique which is being presented in this paper can be applied for modeling a wide class of problems, such as diffusion of pollutants and population models due to the presence of arbitrary functions and free parameters. Besides it can be very useful to generate the testing problems for numerical modeling.\\
 The whole paper comprises of five sections. Section 1 contains a brief introduction and literature survey. The rest of the manuscript is organized as follows: How to construct the exact solutions in the form of a generalized traveling-wave equation is presented in Section 2. Some examples are also given in Section 2 to validate the study. Some generalization and required transformations for the application of the proposed technique to more complex partial differential equations are presented in Section 3. Some special cases of and deductions from the main results of this study are discussed in Section 4. The study is substantiated in Section 5 with graphical analysis.
 
\section{Solutions of generalized RDEs with delay}
 \par The aim is to find exact solutions of (\ref{rd1}) which take the form
 \begin{equation}\label{rd2}
u=U(y), ~~~~ y=t+\int h(x)dx,
 \end{equation}
 which is the generalized traveling-wave equation.
 Substitution of (\ref{rd2}) into (\ref{rd1}) yields
\begin{equation}\label{rd3}
 \begin{split}
&a(x)h^2U''_{yy}+\left(\left[a(x)h\right]'_x-q(x)+p(x)h\right)U'_y\\
&-c(x)g(U)U'_y+b(x)hG\left(U, W\right)U'_y=0,
\end{split}
\end{equation}
where $W = U(y - \tau)$ and $h = h(x).$
The coefficients of the equation are chosen such that they conform to the relations 
\begin{align}
b(x)&=a(x)h(x),\label{rd4}\\
c(x)&=-a(x)h^2(x),\label{rd5}\\
q(x)&=\left[a(x)h(x)\right]'_x+p(x)h(x)+ka(x)h^2(x),\label{rd6}
\end{align}
where $k$ is a constant. The relations reduce (\ref{rd3}) to
\begin{equation}\label{rd7}
U''_{yy}+\left[G\left(U, W\right)+g(U)-k\right]U'_y=0, ~~~ W = U(y - \tau),
\end{equation}
which is a delay ODE.
A standard form for (\ref{rd6}) is
\begin{equation}\label{rd8}
a(x)h'_x+ka(x)h^2+\left[p(x)+a'(x)\right]h-q(x)=0.
\end{equation}
Given the functions $a(x), p(x)$ and $q(x),$ the relation (\ref{rd8}) is the Riccati ODE for $h = h(x).$
\begin{remark}
Solving the Riccati equation (\ref{rd8}) is not necessary for a given $h = h(x).$ Then the derived generalized traveling-wave equation (\ref{rd2}), is said to be the exact solutions to certain RDEs with delay of the form (\ref{rd1}) for which (\ref{rd4}), (\ref{rd5}) and (\ref{rd6}) are true.
\end{remark}
Consideration is given to the Riccati ODE (\ref{rd8}) under two cases, which are degenerate and nondegenerate.\\
{\bf Degenerate case}. For $k = 0,$ the degeneration of Riccati equation (\ref{rd8}) has the general solution which is given by
\begin{equation}\label{rd9}
h(x)=\frac{\int \nu(x)q(x)dx+C_1}{a(x)\nu(x)}, ~~~ \nu(x)=Exp\left(\int \frac{p(x)}{a(x)}dx\right),
\end{equation}
where $C_1$ denotes an arbitrary constant.
\begin{example}
Consider the case where $a = 1$ and $p(x)= q (x) = x.$ Using (\ref{rd9}) with $C_1=0,$ gives $h(x)=1.$ Substitution of the value of $h$ into (\ref{rd2}), (\ref{rd4}) and (\ref{rd5}) yields $y=t+x, b=1$ and $c=-1.$ Consequently, for functions $g(u)$ and $G(u, w)$ which are arbitrary, the nonlinear RDE with delay
$$\left[x-g(u)\right]u_t=u_{xx}+\left[x+G\left(u, w\right)\right]u_x, ~~w=u(x, t-\tau),$$
is solved by
$$u=U(y), ~~~~ y=t+x,$$
where $U(y)$ is determined by the delay ODE 
 \begin{equation}\label{rd12}
U''_{yy}+\left[G\left(U, W\right)+g(U)\right]U'_y=0, ~~~ W = U(y - \tau),
\end{equation}
which is obtained from (\ref{rd7}) by setting $k = 0.$
\end{example}

\begin{example}
Consider the case where $a = q = 1$ and $p(x)=\frac{1}{x}.$ Using (\ref{rd9}) with $C_1=0,$ gives $h(x)=\frac{x}{2}.$ Substitution of the value of $h$ into (\ref{rd2}), (\ref{rd4}) and (\ref{rd5}) yields $y=t+\frac{x^2}{4}, b=\frac{x}{2}$ and $c=-\frac{x^2}{4}.$ Consequently, for functions $g(u)$ and $G(u, w)$ which are arbitrary, the nonlinear RDE with delay

$$\left[4-x^2g(u)\right]u_t=4u_{xx}+\left[\frac{4}{x}+2xG\left(u, w\right)\right]u_x, ~~w=u(x, t-\tau),$$

is solved by

$$u=U(y), ~~~~ y=t+\frac{x^2}{4},$$

 where $U(y)$ is determined by (\ref{rd12}).
\end{example}

\begin{example}
Consider the case where $a = 1, p=0,$ and $q(x)=x.$ Using (\ref{rd9}) with $C_1=0,$ gives $h(x)=\frac{1}{2}x^2.$ Substitution of the value of $h$ into (\ref{rd2}), (\ref{rd4}) and (\ref{rd5}) yields $y=t+\frac{1}{6}x^3, b=\frac{1}{2}x^2$ and $c=-\frac{1}{4}x^4.$ Consequently, for functions $g(u)$ and $G(u, w)$ which are arbitrary, the nonlinear RDE with delay

$$\left[4x-x^4g(u)\right]u_t=4u_{xx}+2x^2G\left(u, w\right)u_x, ~~w=u(x, t-\tau),$$

is solved by

$$u=U(y), ~~~~ y=t+\frac{1}{6}x^3,$$

where $U(y)$ is determined by (\ref{rd12}).
\end{example}

\begin{example}
Consider the case where $p=0,$ and $a(x)=q(x)=x.$ Using (\ref{rd9}) with $C_1=0,$ gives $h(x)=\frac{1}{2}x.$ Substitution of the value of $h$ into (\ref{rd2}), (\ref{rd4}) and (\ref{rd5}) yields $y=t+\frac{1}{4}x^2, b=\frac{1}{2}x$ and $c=-\frac{1}{4}x^2.$ Consequently, for functions $g(u)$ and $G(u, w)$ which are arbitrary, the nonlinear RDE with delay

$$\left[x-\frac{1}{4}x^2g(u)\right]u_t=\left[xu_x\right]_x+\frac{1}{2}xG\left(u, w\right)u_x, ~~w=u(x, t-\tau),$$

is solved by

$$u=U(y), ~~~~ y=t+\frac{1}{4}x^2,$$

where $U(y)$ is determined by (\ref{rd12}).
\end{example}

\begin{example}
Consider the case where two constant coefficients are given as $p(x)=1$ and $q(x)=0,$ while the third coefficient is  given arbitrarily as $a = a(x).$ Using (\ref{rd9}) with $C_1=0,$ gives $h(x)=\frac{x}{a(x)}.$ Substitution of the value of $h$ into (\ref{rd2}), (\ref{rd4}) and (\ref{rd5}) yields $y=t+\int \frac{x}{a(x)} dx, b=x$ and $c=-\frac{x^2}{a(x)}.$ Consequently, for functions $g(u)$ and $G(u, w)$ which are arbitrary, the nonlinear RDE with delay
 \begin{equation}\label{rd10}
\frac{x^2}{a(x)}g(u)u_t+\left[a(x)u_x\right]_x+\left[1+xG\left(u, w\right)\right]u_x=0,
 \end{equation} 
$w=u(x, t-\tau),$ is solved by

$$u=U(y), ~~~~ y=t+\int \frac{x}{a(x)} dx,$$

where $U(y)$ is determined by (\ref{rd12}). Substitution of the function $a(x)=x^n$ into (\ref{rd10}) produces a nonlinear RDE with delay 
$$x^{2-n}g(u)u_t+\left[x^nu_x\right]_x+\left[1+xG\left(u, w\right)\right]u_x=0, ~~w=u(x, t-\tau),$$
 where $n$ represents any number.
\end{example}

{\bf Nondegenerate case}. When $k\neq 0,$ let
\begin{equation}\label{rd11}
h=\frac{1}{k}\frac{{\psi}'_x}{{\psi}}.
\end{equation}
Substituting (\ref{rd11}) into (\ref{rd8}) produces
$$\frac{a(x)}{k}\left(\frac{{\psi}''_{xx}}{{\psi}}-\frac{{\psi}'_x}{{\psi}^2}\right)+ka(x)\left(\frac{1}{k}\frac{{\psi}'_x}{{\psi}}\right)^2+\left[p(x)+a'(x)\right]\frac{1}{k}\frac{{\psi}'_x}{{\psi}}-q(x)=0,$$
which is simplified to obtain linear ODE of second-order
\begin{equation}\label{rd13}
a(x){\psi}''_{xx}+\left[p(x)+a'(x)\right]{\psi}'_x-kq(x){\psi}=0.
\end{equation}
Interested readers in the exact solutions of (\ref{rd13}) for varieties of the functions $a(x), p(x)$ and $q(x),$ are referred to \cite{Polyanin3, Polyanin4}.
\begin{example}
Consider the case where $a = q= 1$ and $p=0.$ The equation (\ref{rd13}) has the general solution which is given by
\begin{equation}\label{rd14}
{\psi}=\begin{cases}
 {\Lambda}_1~ cos(\lambda x) +  {\Lambda}_2 ~sin(\lambda x), & {\rm }\ \phantom{\infty}\text{if}\,\, k =-{\lambda}^2 < 0,\\
  {\Lambda}_1~ cosh(\lambda x) +  {\Lambda}_2~ sinh(\lambda x), & {\rm }\ \phantom{\infty}\text{if}\,\, k = {\lambda}^2 > 0,\\
\end{cases}
\end{equation} 
 where $ {\Lambda}_1$ and $ {\Lambda}_2$ are arbitrary constants. Setting $ {\Lambda}_1=1,  {\Lambda}_2 = 0,$ and $k =-1 ~\left(< 0\right)$ in (\ref{rd14}), and using formula (\ref{rd11}) gives
$$h(x) = tan~ x.$$
Substituting the function $h$ into (\ref{rd4}) and (\ref{rd5}) gives
$$b(x)= tan~ x, c(x)= -{tan}^2 x.$$
Hence, for arbitrary functions $g(u)$ and $G(u, w),$ the nonlinear RDE with delay
 \begin{equation}\label{rd15}
\left[1-{tan}^2 x~ g(u)\right]u_t=u_{xx}+tan~ x~ G\left(u, w\right)u_x,
 \end{equation} 
is solved by
\begin{equation}\label{rd16}
u=U(y), y = t - In~ cos~ x,
\end{equation} 
where $U(y)$ is determined by the delay ODE
\begin{equation}\label{rd17}
U''_{yy}+\left[G\left(U, W\right)+g(U)+1 \right]U'_y=0, ~~~ W = U(z - \tau).
\end{equation}
\end{example}
\section*{Constructing exact solutions when the function {\boldmath $h(x)$} is given}
 Table \ref{rd18} gives the list of other possible ways which do not involve integrating the Riccati equation (\ref{rd8}), for constructing solutions which are exact to equations of the form (\ref{rd1}). Among the functions $h(x), a(x), p(x),$ and $q(x),$ three functions which include $h(x)$ are assumed to be given. The unknown function which remains is then derived from (\ref{rd8}).  The whole nonlinear RDE of the form (\ref{rd1}) is determined by using (\ref{rd4}) and (\ref{rd5}).
\begin{table}[h!]
    \centering
    \caption{Possible ways of deriving the unknown function when out of the functions $h(x), a(x), p(x),$ and $q(x),$ three functions which include $h(x)$ are given. Note that $k$ and $C_2$ are arbitrary constants.}
    \label{rd18}
    \begin{tabular}{p{0.03\linewidth}  p{0.32\linewidth}  p{0.5\linewidth}}
      & & \\
      No. & Given functions & Derived function \\
      \hline
\multirow{2}{*}{1}& $h = h(x), a = a(x),$ & \multirow{2}{*}{$q(x) = \left[ah\right]'_x+ph+kah^2.$}\\
   & $p = p(x).$ & \\
\multirow{2}{*}{2} & $h = h(x), a = a(x),$ & \multirow{2}{*}{$p(x) = \left(q-ah'_x-kah^2\right)/h-a'.$}\\
 & $q = q(x). $ &\\
\multirow{2}{*}{3} & $h = h(x), p = p(x),$ & $a(x) = \frac{\int \mu \left(q/h-p\right)dx + C_2}{ \mu},$ \\
 & $q = q(x).$ &  where $\mu = h\ \mbox{exp}(k\int h dx).$\\
      \hline
    \end{tabular}
\end{table}
\begin{example}\label{rd21}
Illustration is given by using the third way of deriving the unknown function which is described in Table \ref{rd18}. The $h = h(x)$ is  arbitrary, while $C_2 = 0, q=0,$ and $p=-1.$ Two possible cases are being considered.
\begin{itemize}
	\item [(I)] Degenerate case $k = 0.$ It is obtained that $a(x) = \int h dx / h.$ Consequently, for functions $g(u)$ and $G(u, w)$ which are arbitrary, the nonlinear RDE with delay
	 \begin{equation}\label{rd19}
c(x)g(u)u_t=\left[a(x)u_x\right]_x+\left[b(x)G\left(u, w\right)-1\right]u_x, 
 \end{equation} 
 where $a(x) = \int h dx /h, b(x) = \int h dx,$ and $c(x) = -h\int h dx,$ is solved by
 	 \begin{equation}\label{rd20}
 u=U(y), ~y = t + \int h dx.
 \end{equation}
Here, $U(y)$ in (\ref{rd20}) is determined by (\ref{rd12}).

\item [(II)] Nondegenerate case $k \neq 0.$ A new function $f = f(x)$ is introduced by setting $f(x) = \mbox{exp}(k\int h dx).$ Therefore, it is obtained that $a(x) =  \int h f dx /hf.$ Consequently, for arbitrary functions $g(u)$ and $G(u, w),$ the equation (\ref{rd19}), where $a(x) = \int h f dx /hf, b(x) = \int h f dx /f,$ and $c(x) = -h\int h f dx /f,$ is solved by (\ref{rd20}).
Here, $U(y)$ is determined by (\ref{rd7}).
\end{itemize}
\end{example}

\begin{example}
An elucidation is given for the third way of deriving the unknown function which is described in Table \ref{rd18}. Consider Example \ref{rd21} with $h(x) =1/x.$
\par Degenerate case ($k = 0$): For functions $g(u)$ and $G(u, w)$ which are arbitrary, the nonlinear RDE
$$ln\left(1/\sqrt[x]{x}\right)g(u)u_t=\left[x~ln x~u_x\right]_x+\left[ln x~G\left(u, w\right)-1\right]u_x, $$
is solved by
$$ u=U(y), ~y = t + ln~ x,$$
where $U(y)$ is determined by (\ref{rd12}).\\ \\
\par Nondegenerate case ($k \neq 0$): For arbitrary functions $g(u)$ and $G(u, w),$  the equation
$$\frac{1}{kx}g(u)u_t+\left[\frac{x}{k}u_x\right]_x+\left[\frac{1}{k}G\left(u, w\right)-1\right]u_x=0,$$
 is solved by (\ref{rd20}). Here,  $U(y)$ is determined by (\ref{rd7}).
\end{example}
\section{Some transformations and generalization}
The following parameters are assumed to be chosen such that the below relations are satisfied:
 \begin{eqnarray}\label{rd23}
b(x,t)&=&a(x,t)h(x),\nonumber\\
c(x,t)&=&-a(x,t)h^2(x),\\
q(x,t)&=&\left[a(x,t)h(x)\right]'_x+p(x,t)h(x)\nonumber\\
&&+ka(x,t)h^2(x),\nonumber
\end{eqnarray}
where the constant is $k.$
\begin{itemize}
	\item [(I)] {\bf One spatial dimension nonlinear RDEs type with delay}
\end{itemize}
Consider the nonlinear  RDEs with delay which is given as
 \begin{equation}\label{rd22}
 \begin{split}
&\left[q(x,t)+c(x,t)g(u)\right]u_t=\left[a(x,t)u_x\right]_x\\
&+p(x,t)u_x+b(x,t)h(x)G\left(h, u, w, u_x/h\right),
\end{split}
 \end{equation} 
 where the arbitrary function $G\left(h, u, w, \theta \right)$ takes four arguments, $w=u(x, t-\tau)$ and the delay in time is $\tau > 0.$ The equation (\ref{rd22}) has
 \begin{equation} \label{rd24}
 u=U(y), ~y = t + \int h(x) dx,
 \end{equation} 
 as its general solution. Here, $U(y)$ is determined by the delay ODE 
$$U''_{yy}+(g(U)-k)U'_y+G\left(h, U, W, U'_y\right)=0, ~~~ W = U(z - \tau).$$
This can be easily verified by substituting the functions in (\ref{rd24}) into (\ref{rd22}), while taking into account the relations (\ref{rd23}).
\begin{example}
An illustration is given for the degenerate case $k = 0.$  In the third relation of equation (\ref{rd23}), set $a(x, t) = a(x), p(x, t) = 0,$ and $q(x, t) = 1.$ It is obtained that $ b(x,t) = x, c(x,t) = -\frac{x^2}{a(x)}$ and $h(x) = \frac{x}{a(x)}.$ Also in (\ref{rd22}), setting $G\left(h, u, w, \theta \right)=\varphi (u) {\theta}+ h \psi (u,w){\theta}^2$ leads to the deduction that the nonlinear RDE with delay which is of the form
$$\left[1-\frac{x^2}{a(x)}g(u)\right]u_t=\left[a(x,t)u_x\right]_x+x\varphi (u)u_x+ x\psi (u,w)u^2_x,$$
is solved by
 $$u=U(y), ~z = t + \int \frac{x}{a(x)} dx,$$
where $U(y)$ is determined by the delay ODE 
$$U''_{yy}+g(U)U'_y+G\left(h, U, W, U'_y\right)=0.$$
\end{example}
\begin{remark}
For certain functions $a(x, t), p(x, t),$ and $q(x, t),$ exact solutions of the degenerate case ($k = 0$) for the third equation in the relation (\ref{rd23})can be found in \cite{Polyanin5}.
\end{remark}
\begin{itemize}
\item [(II)] {\bf Systems of nonlinear RDEs with delay}
\end{itemize}
Consider the system of nonlinear RDEs with delay which is given as
\begin{equation}\label{rd25}
\begin{split}
 &\left[q(x,t)+c(x,t)g_1(u,v)\right]u_t=\left[a(x,t)u_x\right]_x\\
 &+p(x,t)u_x+b(x,t)h(x)G_1\left(h, u, v, w_1, u_x/h, v_x/h\right),\\
 &\\
 &\left[q(x,t)+c(x,t)g_2(u,v)\right]v_t=\left[a(x,t)v_x\right]_x\\
 &+p(x,t)v_x+b(x,t)h(x)G_2\left(h, u, v, w_2, u_x/h, v_x/h\right),
\end{split}
\end{equation}
where $G_1\left(h, u,v, w_1, {\theta}_1, {\theta}_2 \right),$  $G_2\left(h, u,v, w_2, {\theta}_1, {\theta}_2 \right)$ are arbitrary functions which take six arguments, $w_1=u(x, t-\tau), ~w_2=v(x, t-\tau),$ and the delay in time is $\tau > 0.$ The system of equations (\ref{rd25}) is solved by
 \begin{equation} \label{rd26}
 u=U(y), v=V(y), ~y = t + \int h(x) dx.
 \end{equation} 
Here, $U(y)$ and $V(y)$ are determined by the coupled delay ODE
\begin{align*}
 U''_{yy}+(g_1(U,V)-k)U'_y+G\left(h, U, V, W_1, U'_y, V'_y\right)&=0, \\
V''_{yy}+(g_2(U,V)-k)V'_y+G\left(h, U, V, W_2, U'_y, V'_y\right)&=0,
\end{align*}
where $W_1 = U(y - \tau)$ and $W_2 = V(y - \tau).$
This can be easily verified by substituting the given functions in (\ref{rd26}) into (\ref{rd25}), while taking into account the relations (\ref{rd23}).
\begin{example}
An example is given for the highlight of the degenerate case $k = 0.$ This is done by setting $a(x, t) = 1, ~q(x, t) = 2,$ and $p(x, t) = 0$ in the third relation of equation (\ref{rd23}). It is obtained that $h(x) = 2x, b(x,t) = 2x$ and $c(x,t) = -4x^2.$ By setting  $G_1\left(h, u,v, w_1, {\theta}_1, {\theta}_2 \right)=h f_1(u,v,w) {\theta}_1^2,$  $G_2\left(h, u,v, w_1, {\theta}_1, {\theta}_2 \right)= hf_2(u,v,w) {\theta}_2$ in (\ref{rd25}), the coupled nonlinear reaction-diffusion system with delay
\begin{align*} 
\left[2-4x^2g(u,v)\right]u_t&=u_{xx} + 2xf_1(u, v, w)u^2_x,\\
\left[2-4x^2g(u,v)\right]v_t&=v_{xx} + 4x^2f_2(u, v, w)v_x,
\end{align*}
is solved by
 $$u=U(y), v=V(y), ~y = t + x^2,$$
where $U(y)$ and $V(y)$ are determined by the coupled ODEs with delay
\begin{align*} 
U''_{yy}+g_1(U, V)U'_y+G_1\left(h, U, V, W, U'_y, V'_y\right)&=0,\\
V''_{yy}+g_1(U, V)U'_y+G_2\left(h, U, V, W, U'_y, V'_y\right)&=0.
\end{align*}
\end{example}
\section{Some transformations and deductions}
Special cases of (\ref{rd1}) are being discussed in this section.
\begin{itemize}
\item [(I)] Consider the case $c(x) = p(x) = 0.$ 
\end{itemize}
The nonlinear delay RDE 
 \begin{equation}\label{rd28}
q(x)u_t=\left[a(x)u_x\right]_x + b(x)G\left(u, w\right)u_x, ~~w=u(x, t-\tau),
 \end{equation} 
 is obtained. Some existing results can be obtained as corollaries to (\ref{rd28}) (See e.g, \cite{Polyanin1}). 
\begin{itemize}
\item [(II)] Consider the case $c(x) = 0$ and $G\left(u, w\right) = g(u).$
\end{itemize}
Nonlinear RDE of the form
 \begin{equation}\label{rd29}
q(x)u_t=\left[a(x)u_x\right]_x+p(x)u_x+b(x)g(u)
 \end{equation} 
 is obtained. The equation (\ref{rd29}) is a corollary to our results and it represents the classical local equilibrium case. This vindicates that our results are extension of some existing results where such cases were considered (See e.g, \cite{Polyanin2}).
 
 
\section{Graphical illustration}
Examples are given to show the graphical illustration of the solutions of certain RDEs.
\begin{example}
Consider the RDE
 \begin{equation}\label{rd30}
x^2u_t = u_{xx} - xuu_x.
 \end{equation} 
The interval is $0\leq x \leq 1,$ where the time $t\geq0.$ The initial condition which is the solution at $t=0$ is given as
$$u(x,0)=sin(\pi x).$$
Moreover, the Dirichlet boundary conditions are given as
$$ u(0,t)=0\  \mbox{and}\ \ u(1,t)=0.$$
Figure \ref{rd31} shows the graphical illustration for (\ref{rd30}). It displays how $u$ changes with respect to $x$ at $t=2.$
\begin{figure}
\includegraphics[width=6.0cm ,height=8.0cm]{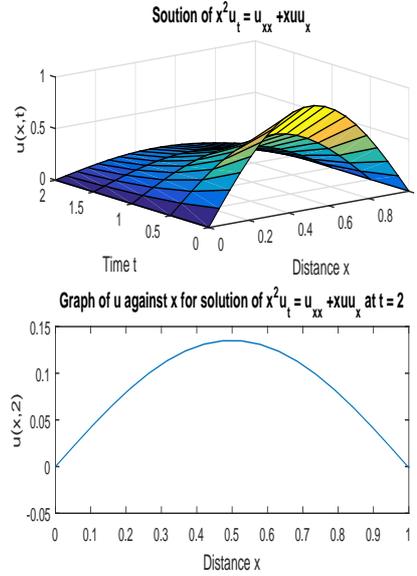}
 \caption{Solutions of $x^2u_t = u_{xx} - xuu_x.$}
 \label{rd31}
\end{figure}
\end{example}
\begin{example}
The convection-diffusion equation is given as
 \begin{equation}\label{rd33}
u_t= u_{xx} + b(x)u_x,
 \end{equation}
 which  has the initial condition as $u(0, x)= \frac{1}{1+(x-5)^2}$ and the Dirichlet boundary conditions $u(t, -\infty) = u(t, +\infty) = 0,$ where 
 $$b(x) = 3\bar{u}(x)^2 - 2\bar{u}(x)$$
 and $\bar{u}$ is implicitly defined by the relation
 $$\frac{1}{\bar{u}}+log \frac{1-\bar{u}}{\bar{u}}=x.$$
 The function $\bar{u}$ is known as an equilibrium solution of
 $$u_t+\left(u^3-u^2\right)_x=u_xx,$$
 where $\bar{u}(-\infty)=1$ and $\bar{u}(+\infty)=0$ (See e.g \cite{Howard}). Figure \ref{rd32} shows the graphical illustration for (\ref{rd33}). A peak is observed to occur at $x=5$ as changes in $u$ with respect to $x$ is being displyed at $t=2.$
 \begin{figure}
\includegraphics[width=6.0cm ,height=8.0cm]{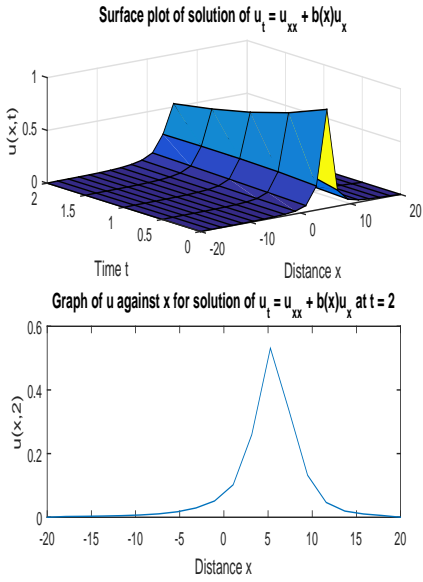}
 \caption{Solutions of $u_t= u_{xx} + b(x)u_x$ (\ref{rd33}).}
 \label{rd32}
\end{figure}
\end{example}
\begin{example}\label{rd40}
Consider a system of RDEs
\begin{equation} \label{rd34}
\begin{split} 
u_t &= u_{xx} + (1 - u - v)u_x,\\
v_t &= v_{xx} + (1 - u - v)v_x,\\
\end{split}
\end{equation}
with the initial conditions
$$\begin{cases} 
u(0, x) = x^2,\\
v(0, x) = x(x - 2),\\
\end{cases}$$
and Cauchy boundary conditions which are specified as
$$\begin{cases} 
u_x(t, 0) = 0; ~u(t, 1) = 1, \\
v(t, 0) = 0; ~v_x (t, 1) = 0.\\
\end{cases}$$

\begin{figure}
\includegraphics[width=6.0cm ,height=8.0cm]{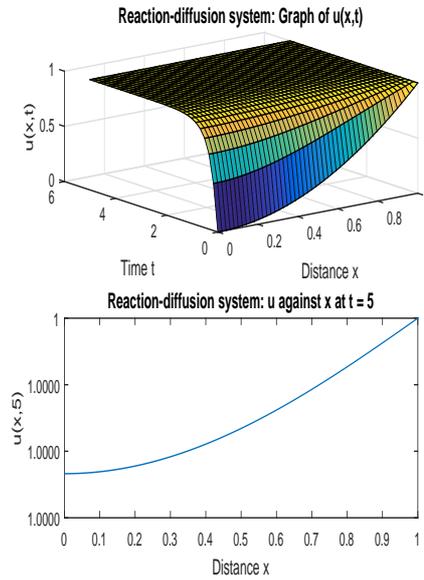}
 \caption{Solutions for $u$ in the system of reaction-diffusion (\ref{rd34}).}
 \label{rd35}
\end{figure}
\begin{figure}
\includegraphics[width=6.0cm ,height=8.0cm]{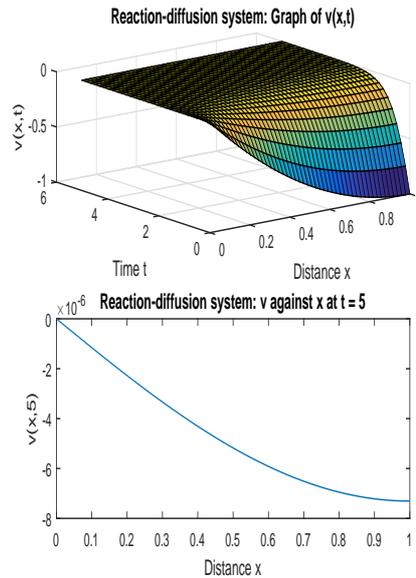}
 \caption{Solutions for $v$ in the system of reaction-diffusion (\ref{rd34}).}
 \label{rd36}
\end{figure}
\end{example}
For the system of reaction-diffusion (\ref{rd34}), the solution for $u$ is displayed in Figure \ref{rd35} while the solution for  $v$ is displayed in Figure \ref{rd36}. Observe that while $u$ has a positive slope, the slope of $v$ is negative.\\ \\
\section{ Conclusion} Recently, exact solutions of  RDEs and reaction-diffusion systems have attracted great attention. In this paper, exact solutions are presented for a most generalized form of  RDE with delay and which are associated with variable coefficients. The presence of arbitrary functions and free parameters in the solutions represents their feasible application in solving certain modeling problems such as diffusion of pollutants and population models, where the population is spatially distributed. It also makes the obtained solutions to be suitable for usage in testing the numerical and approximate analytical methods.  The obtained results also find applications in finding the exact solutions of other form of partial differential equations which are more complex. The examples of delay RDEs with their exact solutions are displayed for elucidation. Graphical illustration of the solutions of some specific RDEs are given.\\

\noindent {\bf Abbreviations}\\
NPDEs: Nonlinear Partial Differential Equations\\
RDEs: Reaction-Diffusion Equations\\ \\

\section*{Declaration of competing interest}
The authors declare that they have no known competing financial interests or personal relationships that could have appeared to
influence the work reported in this paper.

\section*{Acknowledgements} The first author acknowledges with thanks the postdoctoral fellowship and financial support from the DSI-NRF Center of Excellence in Mathematical and Statistical Sciences (CoE-MaSS). Opinions expressed and conclusions arrived are those of the authors and are not necessarily to be attributed to the CoE-MaSS.\\ \\


\footnotesize
\section*{ References}
\begin{thebibliography}{99}
\bibitem{Nisar4} Li Z, Manafian J, Ibrahimov N, Hajar A, Nisar KS, Jamshed W. \emph{Variety interaction between k-lump and k-kink solutions for the generalized Burgers equation with variable coefficients by bilinear analysis}. Results in Physics 2021; 28: 104490.
\bibitem{Nisar5} Nisar KS, Ali J, Mahmood MK, Ahmad D, Ali S. \emph{Hybrid evolutionary pad$\acute{e}$ approximation approach for numerical treatment of nonlinear partial differential equations}. Alexandria Engineering Journal. 2021; 60 (5): 4411-4421.
\bibitem{Kolmogorov} Kolmogorov A, Petrovskii I, Piskunov N. \emph{Study of a diffusion equation that is related to the growth of a quality of matter and its application to a biological problem}. Moscow University Mathematics Bulletin. 1937; 1, 1-26.
\bibitem{Murray}  Murray JD. \emph{Mathematical Biology}. Springer, Berlin; 1989.
 \bibitem{Wang1} Yue X, Wang F,  Hua Q, Qiu XY. \emph{A novel space–time meshless method for nonhomogeneous convection-diffusion equations with variable coefcients}. Applied Mathematics Letters. 2019; 92: 144-150.
 \bibitem{Wang2} Wang F, Wang C, Chen Z. \emph{Local knot method for 2D and 3D convection-diffusion-reaction equations in arbitrary domains}. Applied Mathematics Letters 2020; 105: 106308.
 \bibitem{Nisar1} Zhang Z, Rahman G, Nisar KS, Agarwal RP. \emph{Incorporating convex incidence rate and public awareness program in modelling drinking abuse and novel control strategies with time delay}. Phys. Scr.  2021; 96: 114006.
\bibitem{Nisar2} Sardar M, Khajanchi S, Biswas S. Abdelwahab SF, Nisar KS. \emph{Exploring the dynamics of a tumor-immune interplay with time delay}. Alexandria Engineering Journal. 2021; 60 (5): 4875-4888.
\bibitem{Nisar3} Shaikh A, Tassaddiq A, Nisar KS, Baleanu D. \emph{Analysis of differential equations involving Caputo–Fabrizio fractional operator and its applications to reaction–diffusion equations}. Advances in Difference Equations 2019; 178.
\bibitem{Nisar6} Zada L, Nawaz R, Ahsan S, Nisar KS, Baleanu D. \emph{New iterative approach for the solutions of fractional order inhomogeneous partial differential equations}. AIMS Mathematics. 2021, 6 (2): 1348-1365.
 \bibitem{Aibinu}	Aibinu MO, Thakur SC, Moyo M. \emph {Solving delay differential equations via Sumudu transform}. arXiv:2106.03515v1. 2021.
\bibitem{Polyanin6} Sorokin VG, Vyazmin A, Zhurov AI, Reznik V, Polyanin AD. \emph{The heat and mass transfer modeling with
time delay}. Chemical Engineering Transactions. 2017; 57: 1465-1470.
\bibitem{Jackiewicza} Jackiewicza Z, Zubik-Kowal  B. \emph{Spectral collocation and waveform relaxation methods for nonlinear
delay partial differential equations}. Applied Numerical Mathematics. 2006; 56: 433-443.
\bibitem{Kumbinarasaiah} Kumbinarasaiah S. \emph{Numerical solution of partial differential equations using Laguerre wavelets collocation method}. Int J Manage Technol Eng. 2019; 9 (1): 3635-3639.
\bibitem{Aziz}  Aziz I, Siraj-ul-Islam I, Asif M. \emph{Haar wavelet collocation method for three-dimensional elliptic partial differential equations}. Comput Math Appl. 2017; 73: 2023-2034.
\bibitem{Alderremy}  Alderremy AA, Elzaki TM, Chamekh M. \emph{Modified Adomian decomposition method to solve generalized Emden–Fowler systems for singular IVP}. Math Probl Eng. 2019: 6097095. https://doi.org/10.1155/2019/6097095.
\bibitem{Elzaki} Elzaki TM, Chamekh M. \emph{Solving nonlinear fractional differential equations using a new decomposition method}. Univ J Appl Math Comput. 2018; 6: 27-35.
\bibitem{Al-Shaeer} Al-Shaeer M. \emph{Solutions to nonlinear partial differential equations by Tan-Cot method}. IOSR J Math. 2013; 5 (3): 6-11.
\bibitem{Rajarama}  Rajarama J, Chakraverty S. \emph{Residual power series method for solving time-fractional model of vibration equation of large membranes}. J Appl Comput Mech. 2018; 5:  603-615.
\bibitem{Bilidik}  Bilidik N. \emph{General convergence analysis for the perturbation iteration technique}. Turk J Math Comput Sci. 2017; 6: 1-9.
\bibitem{Jena} Jena RM, Chakraverty S. \emph{A new iterative method based solution for fractional Black-Scholes option pricing equations (BSOPE)}, SN Appl Sci. 2019; 1 (1): 95. https://doi.org/10.1007/s42452-018-0106-8.

\bibitem{Polyanin1}  Polyanin AD. \emph{Generalized traveling-wave solutions of nonlinear reaction-diffusion equations with delay and variable coefficients}. Applied Mathematics Letters. 2019; 90: 49-53.
\bibitem{Polyanin2} Polyanin AD. \emph{Functional separable solutions of nonlinear reaction-diffusion equations with variable coefficients}. Applied Mathematics and Computation. 2019; 347; 282-292.
\bibitem{Polyanin3} Polyanin AD, Zaitsev VF. \emph{Handbook of ordinary differential equations: Exact solutions, methods, and problems}. 2nd ed. Boca Raton: Chapman \& Hall/CRC Press; 2003.
\bibitem{Polyanin4}Polyanin AD, Zaitsev VF. \emph{Handbook of ordinary differential equations: exact solutions, methods, and problems}. Boca Raton: CRC Press; 2018.
\bibitem{Polyanin5} Polyanin AD, Nazaikinskii VE. \emph{Handbook of linear partial differential equations for engineers and scientists}. 2nd ed. Boca Raton: CRC Press; 2016.
\bibitem{Howard} Howard P. \emph{Partial differential equations in Matlab 7.0}. Spring; 2005.\\ https://www.math.tamu.edu/~phoward/m442/pdemat.pdf


\end{thebibliography}
\end{document}